\documentclass{aadebug}

\corr{0309027}{237}

\newtheorem{Def}{Definition}

\begin{document}

\runningheads{Schaubschl\"ager, Kranzlm\"uller, Volkert}
{Event-based Program Analysis with DeWiz}

\title{Event-based Program Analysis with DeWiz}

\author{
Christian~Schaubschl\"ager\addressnum{1},
Dieter~Kranzlm\"uller\addressnum{1},
Jens~Volkert\addressnum{1}
}

\address{1}{
GUP,
Joh. Kepler University Linz,
Altenbergerstr. 69,
A-4040 Linz,
Austria/Europe
schaubschlaeger@gup.uni-linz.ac.at
}

% This information will show up in `Document Properties' in Acrobat Reader
\pdfinfo{
/Title (Event-based Program Analysis with DeWiz)
/Author (Christian Schaubschl\"ager et al.)
}

\begin{abstract}
Due to the increased complexity of parallel and distributed programs, debugging
of them is considered to be the most difficult and time consuming part of the
software lifecycle. Tool support is hence a crucial necessity to hide complexity
from the user. However, most existing tools seem inadequate as soon as the
program under consideration exploits more than a few processors over a long
execution time. This problem is addressed by the novel debugging tool DeWiz
(Debugging Wizard), whose focus lies on scalability. DeWiz has a modular,
scalable architecture, and uses the event graph model as a representation of the
investigated program. DeWiz provides a set of modules, which can be combined to
generate, analyze, and visualize event graph data. Within this processing
pipeline the toolset tries to extract useful information, which is presented to
the user at an arbitrary level of abstraction. Additionally, DeWiz is a
framework, which can be used to easily implement arbitrary
user-defined modules.
\end{abstract}


\section{Introduction}

It is well known, that performance analysis and program debugging, respectively, are
two of the most time consuming and complex parts of the software life-cycle. This
is especially true for parallel or distributed programs, since parallelism and
(inter-process) communication introduce new obstacles which are unknown in
sequential programs, and increase the complexity of the software development
process.

During the past years many program analysis and debugging tools have
been developed, using different approaches to hide the complexity of the analyzed
or debugged program from the user. Due to the (at least) two-dimensional nature
of the analysis data, namely time and space (in terms of processes), some kind of graphical
representation has turned out to be the most useful way to present the analysis
data to the user. Several approaches of graphical representation have been
proposed, most of them visualize a given program execution as a two-dimensional
space-time diagram. There is a broad range of tools in this field, for example
Vampir \cite{Nage96} and Paradyn \cite{Mill94}, just to list two. Some tools use
three dimensional environments like a CAVE to visualize a program execution,
for example as a Time Tunnel as described in \cite{Reed95}.

A characteristic of parallel programs, which is becoming
increasingly important for tool developers, is scalability. With multiprocessor
machines and clusters deploying hundreds or thousands of processors, and grid
infrastructures combining large numbers of distributed resources, scalability of
program analysis tools seems a basic necessity.
An important factor which limits scalability of tools, is the sheer amount of
analysis data. Therefore it is inevitable for any analysis tool to keep
the amount of data presented to the user at manageable sizes. This can be achieved in
two ways: firstly by addressing the data collection phase, i.e. by reducing the
actual amount of collected data. This approach is utilized in Paradyn, where
the amount of collected data is reduced through dynamic
instrumentation \cite{HoMi94}. The underlying idea is to extract only those data
items, that are actually needed for program analysis. This reduces the
total amount of analysis data and thus permits to investigate even large scale
programs.

On the other hand, even with data reduction applied in the collection
phase, the amount of trace data can grow to an enormous size on large scale
programs which utilize a large number of processors ofer a long execution time
which may exceed days, weeks, or even months. This makes it necessary to
focus on scalability also during the data analysis phase. Obviously trace
data must be analyzed in a reasonable time and the results must be presented
to the user in a meaningful way. Abstraction and graphical representation are the
two most important concepts to achieve scalability. An example for such an
abstraction mechanism can be found in EDL, the Event Definition Language
introduced by Bates and Wileden \cite{BaWi83}. EDL uses two essential mechanisms for event
abstraction: filtering and clustering. With filtering, all but a designated
subset of events can be deleted from the original event stream. Clustering means,
that one or more primitive events are gathered together into a higher level
event. EDL has lead to the high-level debugging approach EBBA, Event Based
Behavioural Abstraction \cite{Bat95} and the program behaviour models of
FORMAN \cite{Aug98}. Both
models follow the idea that the behaviour observed in parallel programs
may reveal useful patterns, which can be evaluated during program analysis.
Another, more recent approach of program monitoring is EARL and has
been proposed by Wolf and Mohr in \cite{WoMo98}. EARL stands for Event
Analysis and Recognition Language and it allows to construct target independent
monitoring and analysis tools by writing scripts in the EARL language.

In this
paper we describe the scalable and modular debugging tool DeWiz (Debugging
Wizard), which uses the event graph model to represent a program's execution.
Data analysis and presentation is done by independent modules, which try to
automatically extract useful information. In Section 2 the architecture of DeWiz
is discussed, while in Section 3 we give some examples that show how DeWiz can be
used for program analysis. Finally, an outlook on future work concludes the paper.

\section{Tool Architecture}
The approach of DeWiz stems from our work on the Monitoring and Debugging
environment MAD \cite{KrGr97}. MAD is a collection of software tools for debugging message
passing programs based on the MPI standard \cite{Mpi95}. At the core of this toolset are
the monitoring tool NOPE and the visualization tool ATEMPT. Although originally
developed for message passing programs, the toolset, especially the
monitor NOPE, recently has been extended so that also shared memory codes can be traced.
The motivation for this extension was, that some of todays architectures are
best utilized by using a hybrid MPI/OpenMP programming style \cite{Rab02}.

In the following we will describe the architecture, the theoretical model, as
well as some implementation aspects of DeWiz in more detail.

\subsection{Event Graph}

As mentioned above, in DeWiz program executions as recorded with NOPE
or event streams generated by online monitors
are represented as event graph, which can be defined as follows:

\begin{Def}[Event Graph \cite{Kra00}]
An event graph is a directed graph $G=(E,\rightarrow)$ , where $E$
is the non-empty set of events $e \in E$, while $\rightarrow$ is a relation
connecting events, such that $x \rightarrow y$ means that
there is an edge from event $x$ to event $y$ in $G$ with the "tail" at event
$x$ and the "head" at event $y$.
\end{Def}

The events $e \in E$ of an event graph are the events observed during a program's
execution, like for example send or receive events in message passing programs,
and read or write memory accesses in a shared memory program. In case of NOPE
there is a standard set of events that will be traced, namely (amongst others)
all MPI point-to-point communication events. However, it is easily possible
to specify additional user-defined events to be recorded with NOPE, which adds
great flexibility to the tool.

The relation connecting the events of an event graph is the
{\em happened-before relation},
which is the transitive, irreflexive closure of the union of the
relations $\stackrel{S}{\rightarrow}$ and $\stackrel{C}{\rightarrow}$. It
has been defined as follows:

\begin{Def}[Happened-before relation \cite{Lam78}]
The happened-before relation $\rightarrow$ is defined as\\
\begin{center}
$\rightarrow = (\stackrel{S}{\rightarrow} \cup \stackrel{C}{\rightarrow})^{+}$\\
\end{center}
where $\stackrel{S}{\rightarrow}$ is the sequential order of events
relative to a particular responsible object,
while $\stackrel{C}{\rightarrow}$ is the concurrent order relation connecting
events on arbitrary responsible objects.
\end{Def}

In other words, the relation $\stackrel{S}{\rightarrow}$ defines the sequential
order of events on a particular process, with the meaning that if two events
$e_p^i$ and $e_p^j$ occur on the same process and $e_p^i$ occurs before $e_p^j$
then $e_p^i \stackrel{S}{\rightarrow} e_p^j$.
The concurrent order relation $\stackrel{C}{\rightarrow}$ describes the order
of corresponding events on different processes, which is established by
communication and synchronization. If $e_p^i$ is a send event on process $p$
and $e_q^j$ is the corresponding receive event on process $q$, then
$e_p^i \stackrel{C}{\rightarrow} e_q^j$.

The DeWiz toolset uses the event graph model as its theoretical fundament. The tool
itself consists of three main components, the {\em modules}, the
{\em protocol}, and a {\em framework}, which are required to construct a DeWiz
system for a concrete analysis task.

\subsection{DeWiz System}

A DeWiz system is built by connecting a set of DeWiz modules, which then act
as a kind of event-graph processing pipeline, i.e. the DeWiz modules are
responsible for the actual work in a DeWiz system. This modular approach
has several advantages. It makes the DeWiz system flexible and easily
extensible. Users can utilize existing modules or, if needed, implement their
own modules, hence adding arbitrary functionality to the system.

Basically we distinguish three kinds of modules:

\begin{itemize}
  \item Event graph generation modules
  \item Automatic analysis modules
  \item Data access modules
\end{itemize}

The modules in a DeWiz system communicate with each other using
a specialized protocol,
the DeWiz protocol. This protocol is based upon TCP/IP, which makes it
possible to distribute a DeWiz system across several computers.
Due to this approach, the monitoring and analysis tasks itself can
utilize a potentially large number of resources, e.g. by putting
the analysis tasks on the grid \cite{Foster}.
For example it would be feasible to execute only the
monitoring module on the computer where the monitored application
is running. The monitoring module would then send the collected
events to an analysis module which is executed on some other computer, and
so on. Since analysis or processing of monitored events in general can be
very time-consuming tasks, the distribution of these tasks can speed-up
the analysis process significantly.

As mentioned above we distinguish three types of modules. These will be
described in more detail in the following sections.

\subsubsection{Event Graph Generation Modules}

Event graph generation modules are those who produce the event graph data
stream from a given program execution. This can be done in two ways, either online or
post-mortem. In case of online tracing a DeWiz-Module connects to a running,
instrumented program, collects events which are generated by the online
monitor, and forwards these events to the next module in the DeWiz system.
Currently DeWiz supports online monitors which correspond to the
OMIS Compliant Monitor OCM \cite{WiTr98}. There is also an interface to the
OpenMP Pragma and Region Instrumentator OPARI \cite{MoMa01}.

In case of post-mortem tracing, events are read from tracefiles by a proper
DeWiz module. Currently there is a module for reading tracefiles generated
by NOPE.

\subsubsection{Automatic Analysis Modules}

Automatic analysis modules process an event graph stream and try to extract
useful information like for example communication patterns, or erroneous
behaviour like communication errors. The latter is relatively easy,
for example by simply comparing the message lengths at a send event
and at the corresponding receive event. If the lengths differ, it is an
indication for a possible communication error. A more challenging task
is to try to find communication patterns in an event graph. By applying
pattern-matching algorithms to the event graph, we try to identify patterns
like for example loops. If it is  possible to find any
irregularities in the pattern, this would again be a possible source for
an error in the investigated program.

\subsubsection{Data Access Modules}

At the end of the processing pipeline we have data access modules. Their
purpose is to display the various analysis-results, which were generated by
the predecessing modules, to the user. Depending on the kind of analysis data
a suitable form of visualisation will be chosen. In most cases this will be some
form of graphical representation, for example in form of a space-time diagram of
the event graph.
Figure~\ref{comm_failures} shows a visualization of an example message-passing
event-graph.
On the vertical axes the participating processes are displayed, whereas
the horizontal axes represent the time. The black arrows represent messages
which are sent from one process to another, with the tail of the arrow at
the send event on the source process, and the tip of the arrow at
receive event on the destination process.
The colored arrows indicate possible communication errors; these will be
described in more detail below.

\subsection{The DeWiz Protocol and Framework}

The DeWiz Protocol is used between modules to transport the event graph stream.
For this purpose it is necessary to define data structures which represent the
observed events. In our case the following two data structures have been defined:\\

\begin{center}
  event: $e_p^i = (p,i,type,data)$\\
  \vspace{\baselineskip}
  concurrent order relation: $e_p^i \rightarrow e_q^j = (p,i,q,j)$\\
\end{center}

The variables $p$ and $i$ represent the responsible object (e.g. a process) on which
the event occurred and its sequential order, respectively. The variable $type$
denotes the kind of event, in case of a message passing code a send or a receive
operation for example, or a semaphore lock in a shared memory environment. Currently
only message-passing and shared-memory events are supported, but due to its
flexibility, the event graph can be used to model any kind of software system.
Table~\ref{evt_table} gives a short overview of several possible software
systems, their corresponding event types and event data.
The $data$ variable can be used to store additional information concerning the event,
like for example timestamps or calling parameter of the function call that caused the
event.

\begin{table}
\begin{center}
\begin{tabular}{|p{4cm}|p{3cm}|p{5cm}|}
\hline
   target system & event type & event data \\[0.6cm]
\hline\hline
  parallel/distributed message-passing program &
  send &
  message data, message-length, destination,message-type,data-type,...\\[0.6cm]
\hline
  multi-threaded shared memory program & lock & semaphore, waiting time,...\\[0.6cm]
\hline
  database/transaction system & read record & table, location of table, access time,...\\[0.6cm]
\hline
  file input/output & write & filename, device, buffer size,...\\[0.6cm]
\hline
\end{tabular}
\caption{Example events and event attributes}
\label{evt_table}

\end{center}
\end{table}

The concurrent order relation connects corresponding objects as described above.
In DeWiz we use logical vector clocks as described in \cite{Fidg91} by Fidge to
implement the concurrent order relation.

With the DeWiz Framework it is possible to implement DeWiz modules for any
desired functionality. The Framework is written in the Java
programming language and provides a set of API functions which simplify
the development of user-defined modules, for example by hiding the
DeWiz protocol from the user.

\section{Examples}

\subsection{Overview}

\begin{figure}
\centering
\includegraphics[scale=0.4]{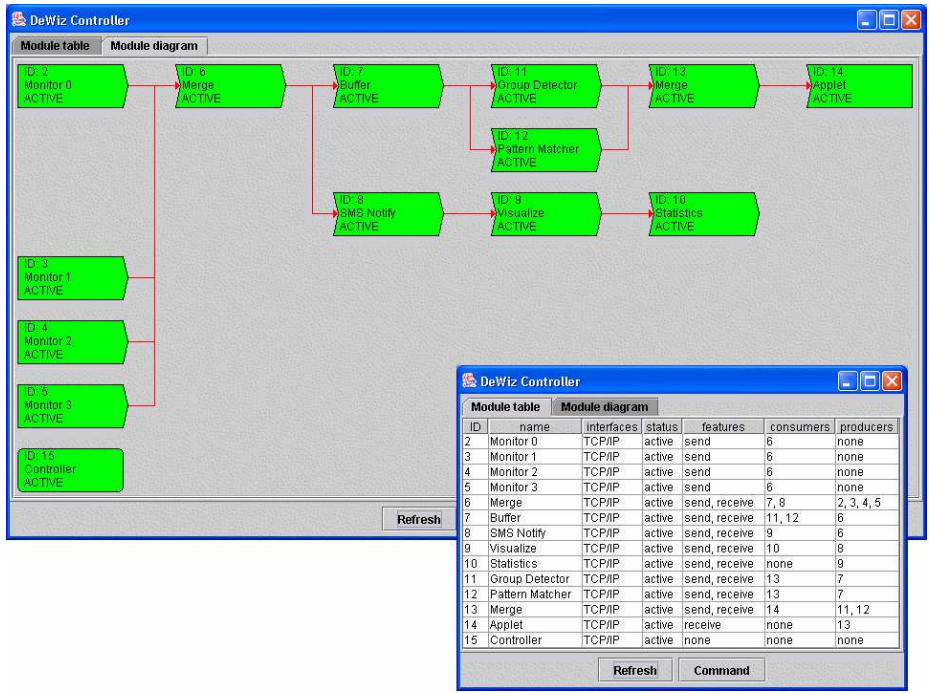}
\caption{An Example DeWiz System}
\label{dewiz}
\end{figure}

In this section we present an example DeWiz system. If the modules for
a concrete analysis task are available, the user may start to construct a
corresponding DeWiz-System. The modules are placed and initialized on arbitrary
networked computing nodes. A dedicated module, the DeWiz Sentinel is used to
control a particular DeWiz System. With a controller interface, available
modules may be arbitrarily interconnected by identifying corresponding input and
output interfaces.
An example for the DeWiz controller interface is shown in Figure~\ref{dewiz}. The smaller
window in front shows the module table, including all registered modules (by id
and name), their available interfaces and status, the implemented features
(send, receive, or none), and the id's of corresponding consumer or producer
modules. The larger background window of Figure~\ref{dewiz} provides the same information
in form of a module diagram.

To use DeWiz in a particular programming environment, dedicated event graph
generation modules have been implemented. As mentioned above, currently there
is a trace-reader modules for NOPE, as well as an interface to OMIS compliant
monitors and an extension to OPARI.

\begin{figure}
\centering
\includegraphics[scale=0.5]{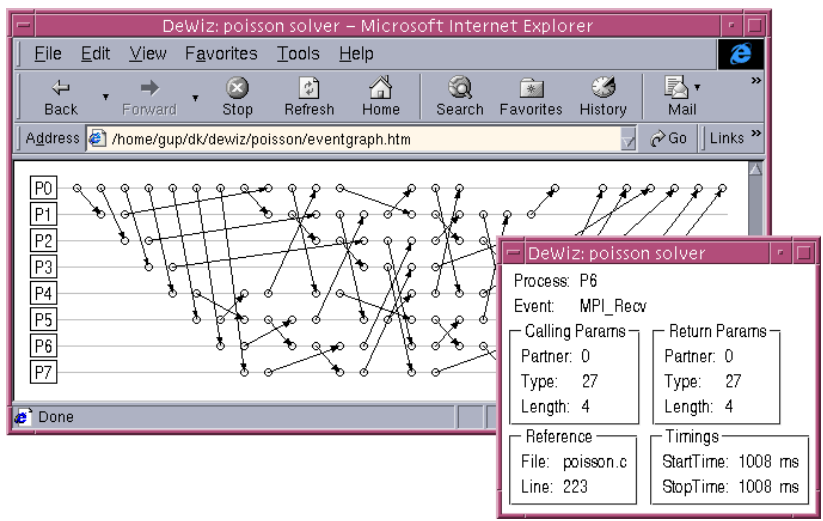}
\caption{Visualization of an event-graph in a Java applet}
\label{iedewiz}
\end{figure}

\begin{figure}
\centering
\includegraphics[scale=0.9]{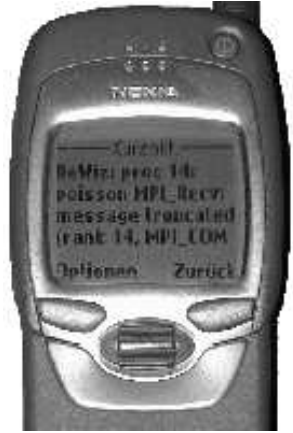}
\caption{DeWiz SMS notifier}
\label{handy}
\end{figure}

Concerning data access modules, DeWiz provides an interface to the analysis tool
ATEMPT (Figure~\ref{comm_failures}), a Java applet to display the event graph stream
in arbitrary web
browsers (Figure~\ref{iedewiz}), and an SMS notifier for critical failures
during program execution (Figure~\ref{handy}).

The analysis functionality already implemented in DeWiz is illustrated with the
following two examples:

\begin{itemize}
  \item Extraction of communication failures
  \item Pattern matching and loop detection
\end{itemize}

\subsection{Communication Failures}

\begin{figure}
\centering
\includegraphics[scale=0.7]{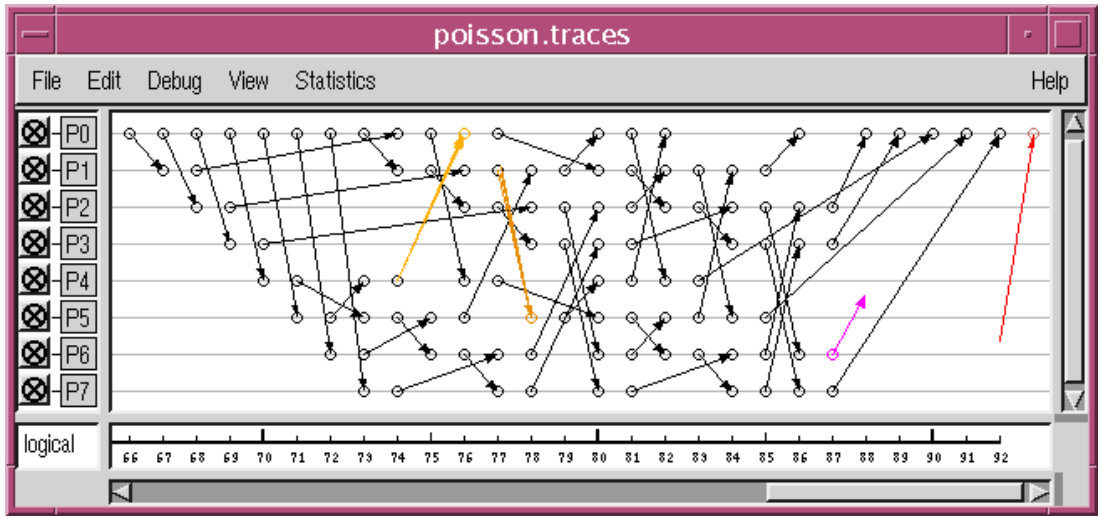}
\caption{Possible communication errors in a message-passing program}
\label{comm_failures}
\end{figure}

Communication failures can be detected by pairwise analysis of communication
events. An example of a possible communication failure is the detection of
different message lengths at a send event and the corresponding receive event.
Though this is not necessarily a communication failure, the default event-graph
visualization module of DeWiz highlights such
send or receive events, respectively, and the user can easily check whether
this is intended or not. Another more obvious example of a communication
failure is the detection of pending send or receive events, which are also
highlighted in the event-graph visualization. Isolated events can originate for
example from a wrong destination address given at a send event. The consequence
would be that the corresponding receive event (in case it is a blocking
receive event) would wait forever for the message, thus blocking the
receiving process forever.
In Figure~\ref{comm_failures}
an example event-graph with several possible communication errors is shown.

\subsection{Pattern Matching - Loop Detection}

\begin{figure}
\centering
\includegraphics{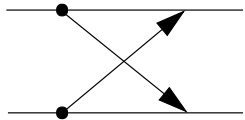}
\caption{Simple exchange}
\label{sim_ex}
\end{figure}

A more complex analysis activity compared to the extraction of communication
failures is pattern matching and loop detection. The goal of the corresponding
DeWiz modules is to identify repeated process interaction patterns in the event
graph. An example event graph is shown in Figure~\ref{sim_ex}. This pattern is called
{\em simple exchange} pattern and can be defined as the event graph\\

\begin{center}

$EX(i,p,q) = (EX_ev(i,p,q),EX_rel(i,p,q))$ with\\
\vspace{\baselineskip}
$EX_ev(i,p,q) = \{e_p^i,e_p^{i+1},e_q^i,e_q^{i+1} \}$ and\\
\vspace{\baselineskip}
$EX_rel(i,p,q)=\{(e_p^i \stackrel{S}{\rightarrow} e_p^{i+1}),
                 (e_q^i \stackrel{S}{\rightarrow} e_q^{i+1}),
                 (e_p^i \stackrel{C}{\rightarrow} e_q^{i+1}),
                 (e_q^i \stackrel{C}{\rightarrow} e_p^{i+1}) \}$
\end{center}

where events $e_p^i, e_p^{i+1}$ occur on
process $p$ and events $e_q^i, e_q^{i+1}$ occur on process $q$ with $p \neq q$.
The existence of this simple pattern in an event graph can easily be verified
within a DeWiz module. More complex
patterns can be specified and provided in a pattern database according to
the needs of users and the characteristics of their programs.

The purpose of detecting patterns in an event-graph is two-fold. Firstly,
if it is possible to detect repeated iterations of a pattern in an event
graph, this knowledge can be used when the event-graph is visualized,
e.g. as space-time diagram. By replacing the possible complicated patterns
with simpler symbols, the complexity of the visual representation of the
event-graph can be reduced greatly, which would give the user a better
overview of the investigated program.

\begin{figure}
\centering
\includegraphics[scale=0.6]{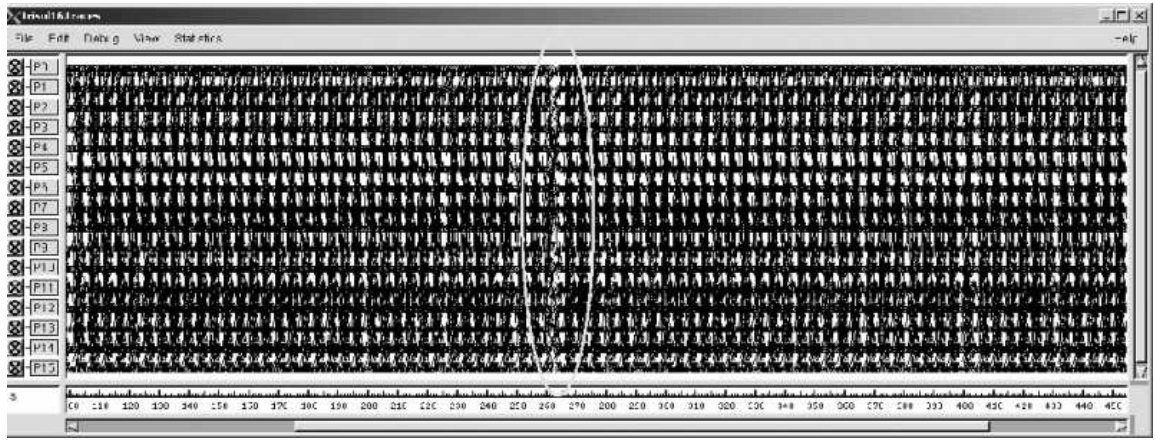}
\caption{Event-graph with iterations of a pattern}
\label{pattern}
\end{figure}

Secondly, the user could specify a communication pattern which is
expected to occur in the investigated program. DeWiz will compare the
given pattern with the event-graph and detect possible deviations, which
could possibly originate from an error in the program. Another example is
the repeated occurrence of any pattern, possibly within a loop. DeWiz will
in a first step detect the pattern, and then check for irregularities in
the sequence of this pattern. Figure~\ref{pattern} illustrates such a situation.
We see a relatively complex event-graph, which is the trace of an
execution of a finite-element message-passing program executed on 16 processes.
Despite its complexity, one can relatively easy see the iterations of a pattern,
as well as a significant irregularity (in the middle of the diagram). Again,
this is an indication for a possible communication error.

\section{Conclusion and Future Work}

Performance analysis and debugging of parallel and distributed programs is a
difficult activity. The problems are further increased, if program executions
with large numbers of processes need to be investigated. For that reason,
scalability of software analysis tools is an important characteristic.

The modular approach of DeWiz provides scalable parallel program analysis by
abstracting the program's behavior as an event graph and distributing the
analysis activities of this graph across existing resources. With this approach,
DeWiz is able to cope with very large amounts of analysis data, while providing
capabilities comparable to existing analysis tools.
The current implementation of DeWiz represents a first proof of concept.
However, for actual application of DeWiz more examinations with real-world
applications are needed. In addition, some more interfaces to existing analysis
tools are required. With the flexible structure of DeWiz and the well-defined
protocol, an interface to an already existing analysis tool can easily be
established. In this way, the analysis tool benefits from the capabilities of
DeWiz and achieves a higher level of scalability.

\bibliography{paper}

\end{document}